\documentclass[aps,prb,twocolumn, titlepage]{revtex4}

\usepackage{graphicx}
\usepackage{color}
\usepackage{dcolumn}
\usepackage{amsfonts}
\usepackage{bm}

\usepackage{epstopdf}

\begin{document}

\title{Phonon spectra of two-dimensional liquid dusty plasmas on a one-dimensional periodic substrate}

\author{W. Li$^1$, D. Huang$^1$, K. Wang$^1$, C. Reichhardt$^2$, C. J. O. Reichhardt$^2$, M. S. Murillo$^3$, and Yan Feng$^1 \ast$}
\affiliation{
$^1$ Center for Soft Condensed Matter Physics and Interdisciplinary Research, School of Physical Science and Technology, Soochow University, Suzhou 215006, China\\
$^2$ Theoretical Division, Los Alamos National Laboratory, Los Alamos, New Mexico 87545, USA\\
$^3$ Department of Computational Mathematics, Science and Engineering, Michigan State University, East Lansing, Michigan 48824, USA\\
$\ast$ E-mail: fengyan@suda.edu.cn}

\date{\today}

\begin{abstract}

We investigate the phonon spectra of two-dimensional liquid dusty plasmas on a one-dimensional periodic substrate using numerical simulations.  The propagation of the waves across the potential wells of the substrate is inhibited due to the confinement of the dust particles by the substrate minima. If the substrate wells are narrow or deep, one-dimensional chains of particles are formed in each minima, and the longitudinal motion of an individual chain dominates the propagation of waves along the potential wells of the substrate. Increasing the width or decreasing the depth of the substrate minima allows the particles to buckle into a zig-zag structure, and the resulting spectra develop two branches, one for sloshing motion and one for breathing motion.  The repulsion between neighboring dust particles produces backward propagation of the sloshing wave for small wave numbers.
\end{abstract}

\maketitle

\section{Introduction}

Dynamical behaviors of an assembly of particles moving over substrates have been widely studied over the past decades in colloidal systems~\cite{Baumgartl:2001,Sarlah:2005,Schmiedeberg:2006,Mikhael:2008,Herrera:2009}, where rich behaviors
such as intriguing phase transitions occur. The interference of two laser beams creates an optical potential in the form of a one-dimensional periodic substrate (1DPS) in colloidal experiments~\cite{Wei:1998}. When the strength of the 1DPS is increased, laser-induced freezing and laser-induced melting appear, as observed in experiments~\cite{Chowdhury:1985, Loudiyi:1992_1, Chakrabarti:1994, Bechinger:2001} and verified in simulations~\cite{Loudiyi:1992_2}. Several laser beams can be combined to generate different substrate geometries, including two-dimensional (2D) periodic substrates~\cite{Bechinger:2001}, quasi-periodic substrates~\cite{Bohlein:2012} and quasi-crystalline substrates~\cite{Su:2017}, and experiments show that these substrates produce abundant phases. Studies of dynamical behavior on these substrates~\cite{Reichhardt:2011,Reichhardt:2012} reveal a variety of phenomena such as intermediate sub-diffusion~\cite{Velarde:2009} as well as pinning and depinning dynamics~\cite{Tierno:2012}. More complex substrate geometries combined with an external force on
the particles are currently being explored in a range of different systems.

A dusty plasma~\cite{Fortov:2005,Morfill:2009,Piel:2010,Bonitz:2010,Thomas:2004} is a partially ionized gas containing micron-sized particles of solid matter~\cite{Feng1:2008}. Under typical experimental conditions, these dust particles are negatively charged to $\approx - 10^4 e$. Due to their high charges,
the dust particles are strongly coupled and unable to move past one another easily, so that the
assembly exhibits liquid-like~\cite{Chan:2007,Feng:2010} and solid-like~\cite{Feng:2008,Hartmann:2014} behavior. The charged dust particles can be levitated and confined by the electric field in the plasma sheath, permitting them to self-organize into a single layer, i.e., a 2D suspension~\cite{Feng:2011} with negligible out-of-plane motion~\cite{Qiao:2003,Donko1:2004}. Within this single layer, the interaction between dust particles is a repulsive Yukawa potential~\cite{Konopka:2000}, resulting from the shielding effects of the free electrons and ions. Through video microscopy, the 2D suspension can be imaged at the scale of individual particles, whose positions and motion can be tracked from frame to frame. Langevin dynamical simulations have been widely used to study the behaviors of 2D dusty plasmas ~\cite{Feng:2017,Kong:2016}.

In dusty plasma experiments, the phonon spectra can be directly calculated based on the thermal motion of the dust particles. The results can be compared to theoretical predictions for the dispersion relations of a 2D dusty plasma crystal lattice~\cite{Wang:2001,Kalman:2004}. Over the past decades, the phonon spectra have been obtained in both experiments~\cite{Couedel:2010,Wang:2004,Melzer:2003,Zhdanov:2003,Nosenko:2006} and simulations~\cite{Peeters:1987,Nunomura:2002,Murillo:2003}, including the compressional (longitudinal) mode and the shear (transverse) mode, and the results are consistent with the theoretical dispersion relations. In addition, the phonon spectra or dispersion relations of a 1D chain or a ring of dusty plasma have been investigated experimentally~\cite{Liu:2003,Misawa:2001,Sheridan:2010,Sheridan:2016,Sheridan:2017} and theoretically~\cite{Melandso:1996,Sheridan:2009}. Similar studies have also been performed using two chains or rings as well as mixed Yukawa particles~\cite{Tkachenko:2011,Ferreria:2008,Piacente:2004,Zhang:2017}.
We are, however, unaware of 
any previous studies of the phonon spectra for 2D dusty plasmas on substrates.

In the dusty plasma experiments of Refs.~\cite{Jiang:2009,Li:2009,Li:2010}, a modified electrode was used
to generate a 1D substrate for studying the dynamics and transport response of dusty plasmas.
It may be possible in the future to use
laser-generated substrates
of the type employed in colloidal experiments~\cite{Baumgartl:2001}
for dusty plasma experiments.
The colloidal system is overdamped, but the dusty plasma is underdamped.
Thus, inertial effects in the motion of the dusty plasma particles can generate
rich behavior not found for colloidal particles.

Here, we explore the impact of different types of 1DPS on the collective modes of 2D dusty plasmas. In Sec. II, we briefly describe our Langevin dynamical simulation method. In Sec. III, we present the phonon spectra of the 2D dusty plasmas on varied types of 1DPS. We summarize our work in Sec. IV.

\section{Simulation method}

Dusty plasmas are traditionally characterized using two dimensionless parameters, the coupling parameter $\Gamma$ and the screening parameter $\kappa$. Here, $\Gamma = Q^2/(4 \pi \epsilon_0 a k_B T)$ can be regarded as the inverse temperature, where $Q$ is the particle charge, $T$ is the kinetic temperature of particles, and $a = 1/\sqrt{\pi n}$ is the Wigner-Seitz radius for an areal number density of $n$. The screening parameter $\kappa = a / \lambda_D$ indicates the length scale of the space occupied by one dust particle over the Debye screening length $\lambda_D$. To normalize the length,
in addition to the value of $a$, we
use the average distance between
nearest neighbors, called the lattice constant $b$. For the 2D triangular lattice we consider here, $b = 1.9046a$.

To obtain the dynamics of 2D dusty plasmas on a 1D periodic substrate, we perform Langevin dynamical simulations. We numerically integrate the equation of motion
\begin{equation}\label{LDE}
{	m \ddot{\bf r}_i = -\nabla \Sigma \phi_{ij} - \nu m\dot{\bf r}_i + \xi_i(t)+{\bf F}^{S}_i }
\end{equation}
for 1024 dust particles, confined in a rectangular box with
dimensions $61.1 a \times 52.9 a$, using
periodic boundary conditions. The four terms on the right-hand side of Eq.~(\ref{LDE}) are the particle-particle interaction $-\nabla \Sigma \phi_{ij}$, the frictional drag $- \nu m\dot{\bf r}_i$, the Langevin random kicks $\xi_i(t)$, and the force ${\bf F}^{S}_i$ due to the 1D periodic substrate, respectively. The inter-particle interaction has the form of
a Yukawa potential, $\phi_{ij} = Q^2 {\rm exp}(-r_{ij} / \lambda_D) / 4 \pi \epsilon_0 r_{ij}$, where $r_{ij}$ is the distance between particles $i$ and $j$. The force ${\bf F}^{S}_i$ from the 1D periodic substrate is
given by
\begin{equation}\label{FS}
{	{\bf F}^{S}_i = - \frac {\partial U(x)}{\partial x} = (2\pi U_0/w) \sin (2\pi x/w) \hat{\bf x}, }
\end{equation}
where $U(x) = U_0 \cos(2\pi x/w)$ is an array of potential wells parallel to the $y$ axis.
Here, $U_0$ is the depth of each potential well (or the substrate depth) in
units of $E_0=Q^2/4\pi\epsilon_0 b$ and $w$ is the width of the potential wells (or the substrate period) in
units of $b$.

To obtain a 2D Yukawa liquid, we set the coupling parameter ${\Gamma = 200}$ and the screening parameter ${\kappa = 2}$. This is above the melting point of ${\Gamma=396}$ for the 2D Yukawa system of ${\kappa = 2}$ obtained using the bond-angular order parameter \cite{Hartmann1:2005}. As justified in Ref.~\cite{Liu1:2005}, our integration time step is $0.037{\omega}^{-1}_{pd}$, where ${\omega}_{pd} = (Q^2/2\pi\varepsilon_0 m a^3)^{1/2}$ is the nominal dusty plasma frequency. We take the frictional damping from the gas to be $ \nu / {\omega}_{pd} = 0.027$, comparable to typical experimental conditions~\cite{Feng:2011}. The expression of force ${\bf F}^{S}_i$ in Eq.~(\ref{FS}) has two parameters, the substrate period $w$ and depth $U_0$. Due to the periodic boundary conditions, the substrate period $w$ is
limited to values that produce
integer numbers of potential wells within the simulation box. Here, we choose $w = 1.002b$ and $w = 2.004b$, corresponding to $N_w=32$ and $N_w=16$ potential wells,
respectively. For the substrate depth, we specify three values of $U_0 = 0.25 E_0$, $U_0 = 0.5 E_0$, and $U_0 = E_0$,
in units of $E_0=Q^2/4\pi\epsilon_0 b$.

For each simulation run, after the system has reached a steady state, we
analyze the particle trajectories and velocities
over a period of $10^6$ simulation time steps.
Other details of our simulation are the same as those described in \cite{Feng:2016}. We also performed a few test runs of a larger system containing 4096 dust particles, and found no substantial differences in the
phonon spectra.

We calculate the phonon spectra of our simulated Yukawa liquid using the Fourier transformations of longitudinal and transverse current autocorrelation functions. The current autocorrelation functions are defined as~\cite{Ohta:2000,Liu2:2003}
\begin{equation}\label{LCA}
{	C_{L}({\bf k},t)= \frac 1N \langle [ {\bf k} \cdot {\bf j}({\bf k},t)][{\bf k} \cdot {\bf j}(-{\bf k},0)] \rangle ,}
\end{equation}
for the longitudinal mode and
\begin{equation}\label{TCA}
{	C_{T}({\bf k},t)= \frac {1}{2N} \langle [ {\bf k} \times {\bf j}({\bf k},t)]\cdot [{\bf k} \times {\bf j}(-{\bf k},0)] \rangle ,}
\end{equation}
for the transverse mode, where $ {\bf k} $ is the wave vector and $N$ is the number of particles. Here, ${\bf j}({\bf k},t) = \sum_{j=1}^N {\bf v}_j(t) \exp [i{\bf k}\cdot {\bf r}_j(t)]$ is the vector current of all simulated particles for a given wave vector ${\bf k}$, where ${\bf v}_j(t)$ and ${\bf r}_j(t)$ are the velocity and position of the $j$th particle, respectively. Finally, the phonon spectra can be obtained using
\begin{equation}\label{LTP}
{	\tilde{C}_{L,T}({\bf k},\omega) = \int_0^\infty e^{-i\omega t}C_{L,T}({\bf k},t)\mathrm{d}t}.
\end{equation}
Here, $\tilde{C}_{L}({\bf k},\omega)$ and $\tilde{C}_{T}({\bf k},\omega)$ correspond to the longitudinal and transverse wave spectra, respectively. Due to the
1D substrate, the phonon spectra of our simulated Yukawa liquid in different wave vector directions should be completely different. We focus on four types of phonon spectra labeled as $\tilde{C}_{L}({\bf k}_x,\omega)$, $\tilde{C}_{T}({\bf k}_x,\omega)$, $\tilde{C}_{L}({\bf k}_y,\omega)$, and $\tilde{C}_{T}({\bf k}_y,\omega)$, corresponding to the longitudinal and transverse spectra when the wave vector directions are
along the $x$ and $y$ axes, respectively.

\section{Results and discussions}

\subsection{Lattice structure with different periodic substrates}

\begin{figure}[htb]
	\centering
        	\includegraphics{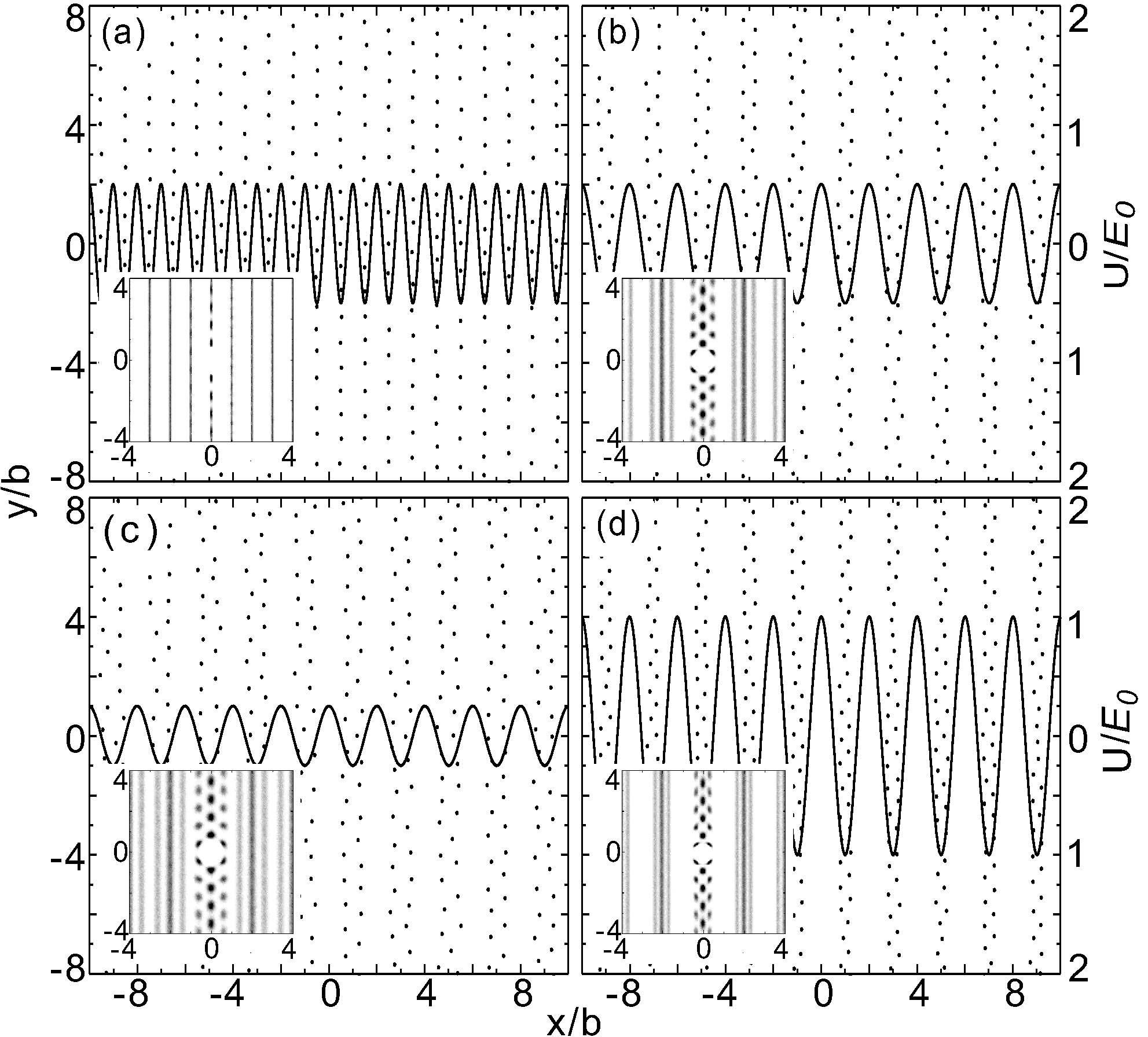}
	\caption{\label{fig:Distribution}
          Snapshots of the dust particle positions (black dots) along with an illustration of the width and depth of the 1D substrate potential.
          (a) $U_0 = 0.5 E_0$ and $w = 1.002 b$. The particles form 1D chains at the bottom of each potential well of the substrate.
          (b) $U_0 = 0.5 E_0$ and $w = 2.004 b$. For this value of $w$, the 1D chain of particles buckles into a zigzag shape.
          (c) $U_0 = 0.25 E_0$ and $w = 2.004b$.
          (d) $U_0 = E_0$ and $w = 2.004 b$. As $U_0$ increases, the zigzag shape narrows. For each panel, the inset on the bottom left corner is the corresponding 2D distribution function~\cite{Loudiyi:1992_1} $g(x,y)$ of the total simulated area.
        }
\end{figure}

Figure~\ref{fig:Distribution} shows the positions of the particles on periodic substrates with different values of $U_0$ and $w$, along with
a plot of the substrate potential.  When the substrate period $w$ is small, the particles are closely confined within each potential well to form 1D chains, as illustrated in Fig.~\ref{fig:Distribution}(a) for $w=1.002b$ and $U_0=0.5 E_0$. If the substrate period is increased but the substrate depth remains unchanged, the particles buckle into a zig-zag chain within each potential well to reduce the interparticle interactions, as shown in Fig.~\ref{fig:Distribution}(b) for $w=2.004b$ and $U_0=0.5 E_0$.  Figure~\ref{fig:Distribution}(c), with $w=2.004b$ and a weaker $U_0=0.25 E_0$, indicates that the width of the zigzag increases if the substrate depth is reduced, while in Fig.~\ref{fig:Distribution}(d) at $w=2.004b$ and a larger $U_0=E_0$, the zigzag narrows and becomes nearly one-dimensional in some regions when the substrate strength is increased. Our calculated 2D distribution functions~\cite{Loudiyi:1992_1} $g(x,y)$ in Fig.~1 also verify the observations above.

\subsection{Wave spectra with different widths of potential wells $w$}

\begin{figure}[htb]
    \centering
    \includegraphics{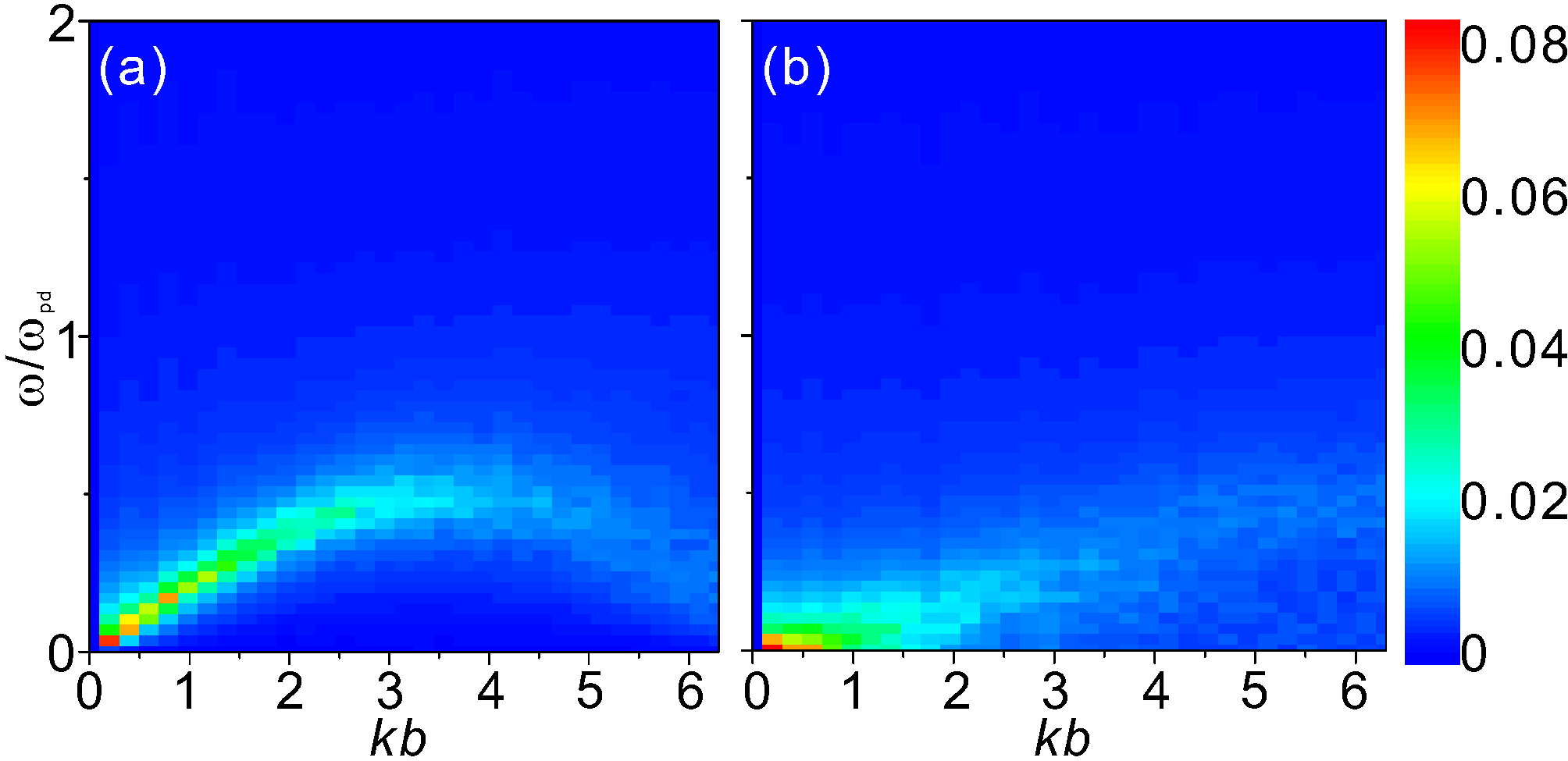}
    \caption{\label{fig:NoSub}
      Longitudinal (a) and transverse (b) phonon spectra of our simulated 2D Yukawa liquid as functions of frequency $\omega/\omega_{pd}$ and wave number $kb$ in the absence of a substrate at $\Gamma = 200$ , $\kappa = 2.0$, and $\nu = 0.027 \omega_{pd}$.
    }
\end{figure}

For comparison, we first compute the phonon spectra $\tilde{C}_{L}({\bf k}_x,\omega)$ and $\tilde{C}_{T}({\bf k}_x,\omega)$ of the 2D Yukawa liquid without a substrate. In Fig.~\ref{fig:NoSub}, we illustrate the longitudinal and transverse phonon spectra of our simulated liquid dusty plasma at the same conditions without any substrate, plotted as height fields as a function of frequency $\omega/\omega_{pd}$ and wave number $kb$. The shape of each spectrum agrees well with previous predictions~\cite{Feng:2014}. Note that, in Fig.~\ref{fig:NoSub}, we only
illustrate the wave vector in the $x$ direction. Since the simulated Yukawa liquid without a substrate is isotropic, we verify that the calculated $\tilde{C}_{L}({\bf k}_x,\omega)$ is nearly the same as $\tilde{C}_{L}({\bf k}_y,\omega)$, and similarly
that
$\tilde{C}_{T}({\bf k}_x,\omega)$ is almost identical to $\tilde{C}_{T}({\bf k}_y,\omega)$.

\begin{figure}[htb]
	\centering
        	\includegraphics{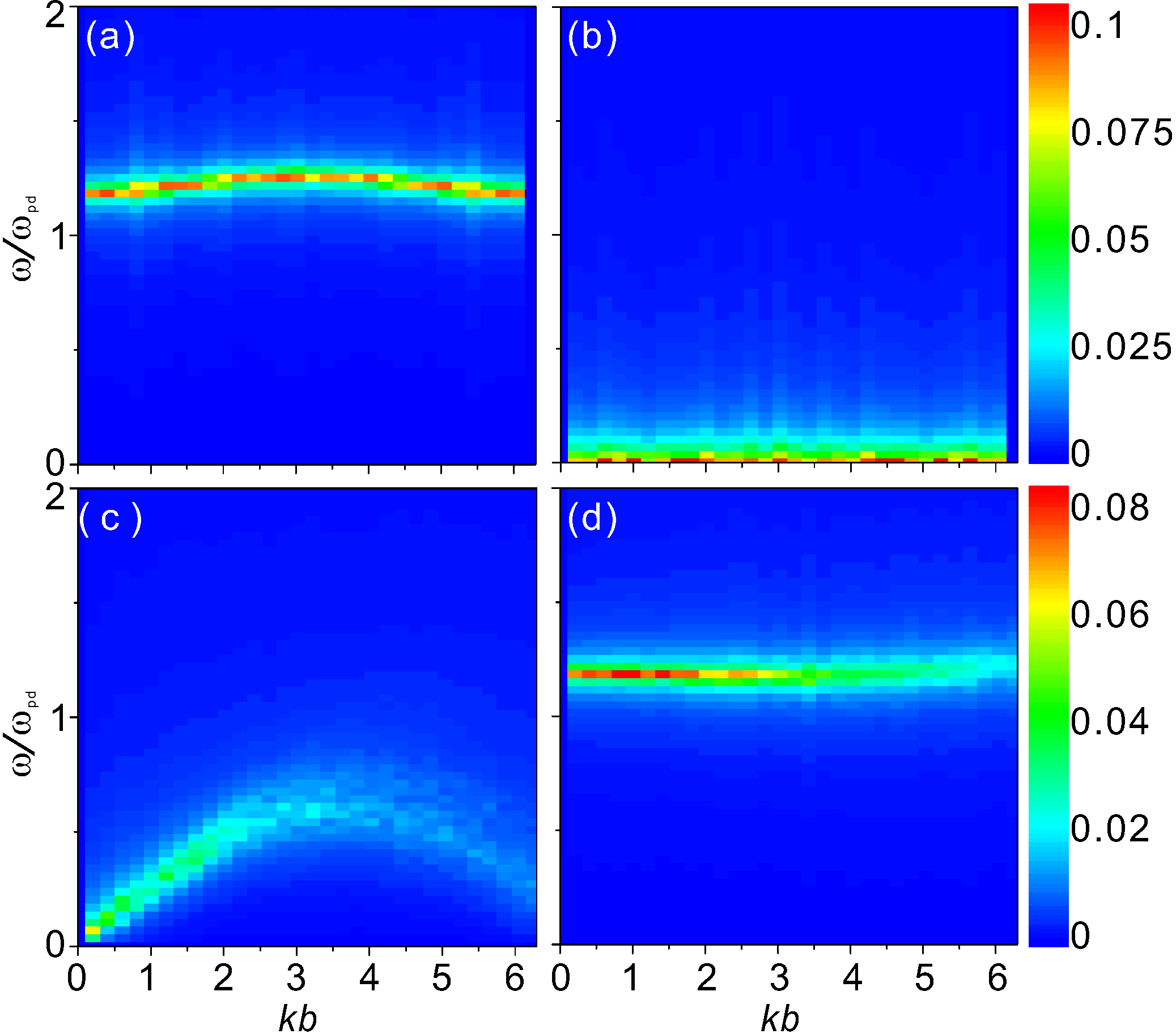}
	\caption{\label{fig:Wid1}
          The longitudinal and transverse phonon spectra $\tilde{C}_{L}({\bf k}_x,\omega)$ (a), $\tilde{C}_{T}({\bf k}_x,\omega)$ (b), $\tilde{C}_{L}({\bf k}_y,\omega)$ (c), and $\tilde{C}_{T}({\bf k}_y,\omega)$ (d) as functions of $\omega/\omega_{pd}$ and $kb$ for a periodic substrate with $U_0=0.5E_0$ and $w=1.002b$.
        }
\end{figure}

Figure~\ref{fig:Wid1} shows the phonon spectra of $\tilde{C}_{L}({\bf k}_x,\omega)$, $\tilde{C}_{T}({\bf k}_x,\omega)$, $\tilde{C}_{L}({\bf k}_y,\omega)$, and $\tilde{C}_{T}({\bf k}_y,\omega)$ in the presence of a periodic substrate
described by
Eq.~(\ref{FS}) with $U_0 = 0.5 E_0$ and $w = 1.002b$. For the $\tilde{C}_{L}({\bf k}_x,\omega)$ and $\tilde{C}_{T}({\bf k}_y,\omega)$ spectra in Figs.~\ref{fig:Wid1}(a) and (d), the flat slope indicates that the group velocity is nearly zero, meaning that the wave cannot propagate. These two spectra correspond to the motion of dust in the $x$ direction, which is completely constrained due to the substrate. The $x$ direction motion of dust particles in different potential wells is very nearly uncorrelated, and this prevents the propagation of waves along the $x$ direction. In these spectra, the frequency is nearly constant at $\omega/\omega_{pd}=1.194$, corresponding to the oscillation frequency of about $1.193 \omega_{pd}$ for a single dust particle within a potential well of depth $U_0 = 0.5 E_0$ and width $w = 1.002b$. Note that the slight difference between the constant frequency in Figs.~\ref{fig:Wid1}(a, d) and the calculated single dust particle oscillation frequency may come from the interparticle interaction and the screening effects of the Yukawa system.

The $\tilde{C}_{T}({\bf k}_x,\omega)$ and $\tilde{C}_{L}({\bf k}_y,\omega)$ spectra in Fig.~\ref{fig:Wid1}(b) and (c) correspond to the motion of dust particles in the $y$ direction, which is unconstrained by the substrate. There is only one chain of dust particles in each potential well, as shown in Fig.~\ref{fig:Distribution}(a), and the motion of dust particles in different potential wells are not correlated at all. As a result, it is not possible for the transverse wave $\tilde{C}_{T}({\bf k}_x,\omega)$ to propagate, and the energy in $\tilde{C}_{T}({\bf k}_x,\omega)$ is distributed around zero frequency,
as shown in Fig.~\ref{fig:Wid1}(b). The $\tilde{C}_{L}({\bf k}_y,\omega)$ spectra in Fig.~\ref{fig:Wid1}(c) is dominated by the longitudinal wave of the single chain within each potential well. Since there is no constraint on the $y$ direction motion, the $\tilde{C}_{L}({\bf k}_y,\omega)$ spectrum in the presence of a substrate is very similar to the substrate-free longitudinal spectrum in Fig.~\ref{fig:NoSub}(a). The sound speed is slightly larger when there is a substrate, as shown in Fig.~\ref{fig:Wid1}(c), because the $y$ direction spacing of the dust particles within each potential well is reduced due to the compression of the substrate, as illustrated in Fig.~\ref{fig:Distribution}(a), giving a stronger interparticle interaction force~\cite{Tkachenko:2011}.

\begin{figure}[htb]
	\centering
        	\includegraphics{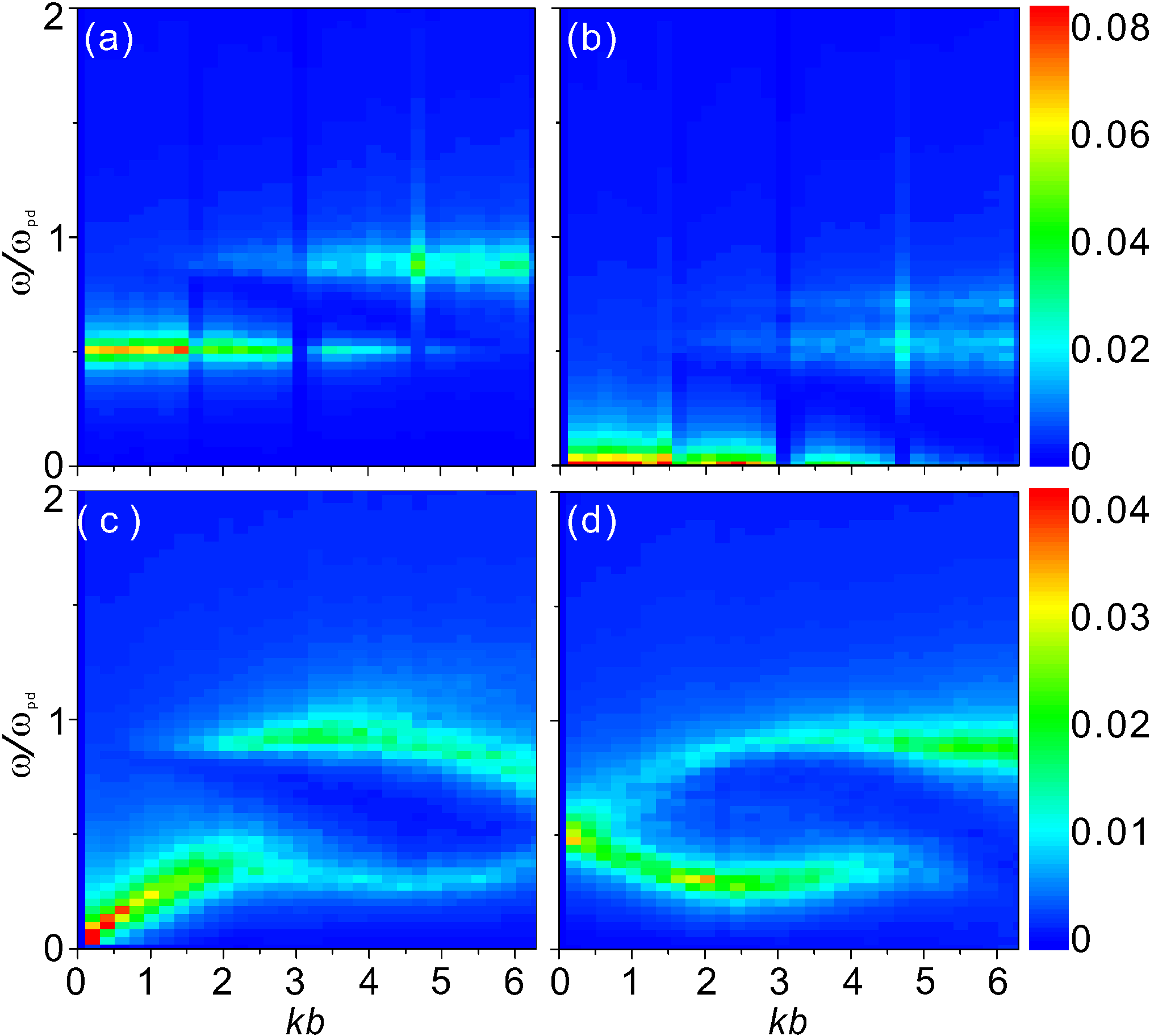}
	\caption{\label{fig:Wid2}
          The longitudinal and transverse phonon spectra $\tilde{C}_{L}({\bf k}_x,\omega)$ (a), $\tilde{C}_{T}({\bf k}_x,\omega)$ (b), $\tilde{C}_{L}({\bf k}_y,\omega)$ (c), and $\tilde{C}_{T}({\bf k}_y,\omega)$ (d) as functions of $\omega/\omega_{pd}$ and $kb$ for a periodic substrate with $U_0 = 0.5 E_0$ and $w = 2.004b$.
          }
\end{figure}

We next consider a substrate with the same depth, $U_0=0.5 E_0$, but with wider wells, $w=2.004b$, and plot the resulting phonon spectra in Fig.~\ref{fig:Wid2}.
All four spectra in Fig.~\ref{fig:Wid2} have two branches due to the zigzag structure that
produces the equivalent of
two chains within each potential well, as shown in Fig.~\ref{fig:Distribution}(b). The $\tilde{C}_{L}({\bf k}_x,\omega)$ and $\tilde{C}_{T}({\bf k}_y,\omega)$ spectra of Figs.~\ref{fig:Wid2}(a) and (d) correspond to the motion of dust particles in the $x$ direction, which is constrained by the substrate. The lower and higher branches roughly correspond to the sloshing \cite{Melzer:2003} and breathing motion of dust particles within each potential well, and we refer to these as the sloshing and breathing branches, respectively.
When the wavenumber is nearly zero, both $\tilde{C}_{L}({\bf k}_x,\omega)$ and $\tilde{C}_{T}({\bf k}_y,\omega)$ have the same frequency of $\omega/\omega_{pd}=0.51$, which is very close to the estimated sloshing mode oscillation frequency of $0.59 \omega_{pd}$ for the two combined dust particles within this potential well.

In Fig.~\ref{fig:Wid2}(d), the wave speed is negative for the sloshing branch when the wavenumber $kb < 2$, which means that this wave propagates backward. The mechanism of the backward motion is the pure repulsion between dust particles, which is the same as in the transverse wave of the 1D Yukawa chain~\cite{Liu:2003}.
When $kb = 0$ in the sloshing mode, all of the dust particles move together, and the restoring force is provided entirely by the potential well of the substrate.
When the wavenumber is slightly larger than zero, neighboring dust particles have slightly different $x$ direction coordinates, so the repulsive force between dust particles in the $x$ direction partially cancels the restoring force. In other words, the restoring force is reduced when the wavenumber is larger than zero, reducing the frequency and resulting in a backward wave~\cite{Liu:2003,Nunomura:2002}.

The $\tilde{C}_{T}({\bf k}_x,\omega)$ and $\tilde{C}_{L}({\bf k}_y,\omega)$ spectra in Figs.~\ref{fig:Wid2}(b) and (c) represent the motion of dust particles in the $y$ direction, which is unconstrained by the substrate. The lower branch of Fig.~\ref{fig:Wid2}(b) is similar to that of Fig.~\ref{fig:Wid1}(b), and corresponds to the uncorrelated motion of dust particles in different potential wells. The upper branch in Fig.~\ref{fig:Wid2}(b) arises from the correlated motion of the
two chains formed by the zigzag structure within one potential well, as shown in Fig.~\ref{fig:Distribution}(b). Due to the interparticle interactions, the frequency of the upper branch of $\tilde{C}_{T}({\bf k}_x,\omega)$ is not zero, but the group velocity is zero, suggesting that the wave cannot propagate due to the confinement by the potential well. The $\tilde{C}_{L}({\bf k}_y,\omega)$ spectrum in Fig.~\ref{fig:Wid2}(c) splits into two branches above $kb \approx 2$. The lower branch in Fig.~\ref{fig:Wid2}(c) is similar to Fig.~\ref{fig:Wid1}(c), and corresponds to a longitudinal wave of all the dust particles within one potential well at a large length scale. The upper branch of Fig.~\ref{fig:Wid2}(c) is produced by the relative motion of dust particles in the $y$ direction between two chains within one potential well of the substrate, as illustrated in Fig.~\ref{fig:Distribution}(b).

Note that, in our calculated $\tilde{C}_{L}({\bf k}_x,\omega)$ and $\tilde{C}_{T}({\bf k}_x,\omega)$, we observe a few gaps when $kb \approx \pi$. This occurs due to the modulation of the dust particle density by the periodic substrate. In the $x$ direction, the length of the first Brillouin zone~\cite{Born:1954} due to the 1DPS is simply $kb = 2 \pi / w = 3.135$. The spectral information at $kb = 3.135$ is exactly the same as that at $kb = 0$, as shown in Figs.~\ref{fig:Wid2}(a) and (b), where the spectral weight is zero in both cases. Since the motion in the $y$ direction is unconstrained, the $\tilde{C}_{L}({\bf k}_y,\omega)$ and $\tilde{C}_{T}({\bf k}_y,\omega)$ spectra do not show a similar gap feature at all.

\subsection{Wave spectra with different depths of potential wells $U_0$ }

\begin{figure}[htb]
    \centering
        \includegraphics{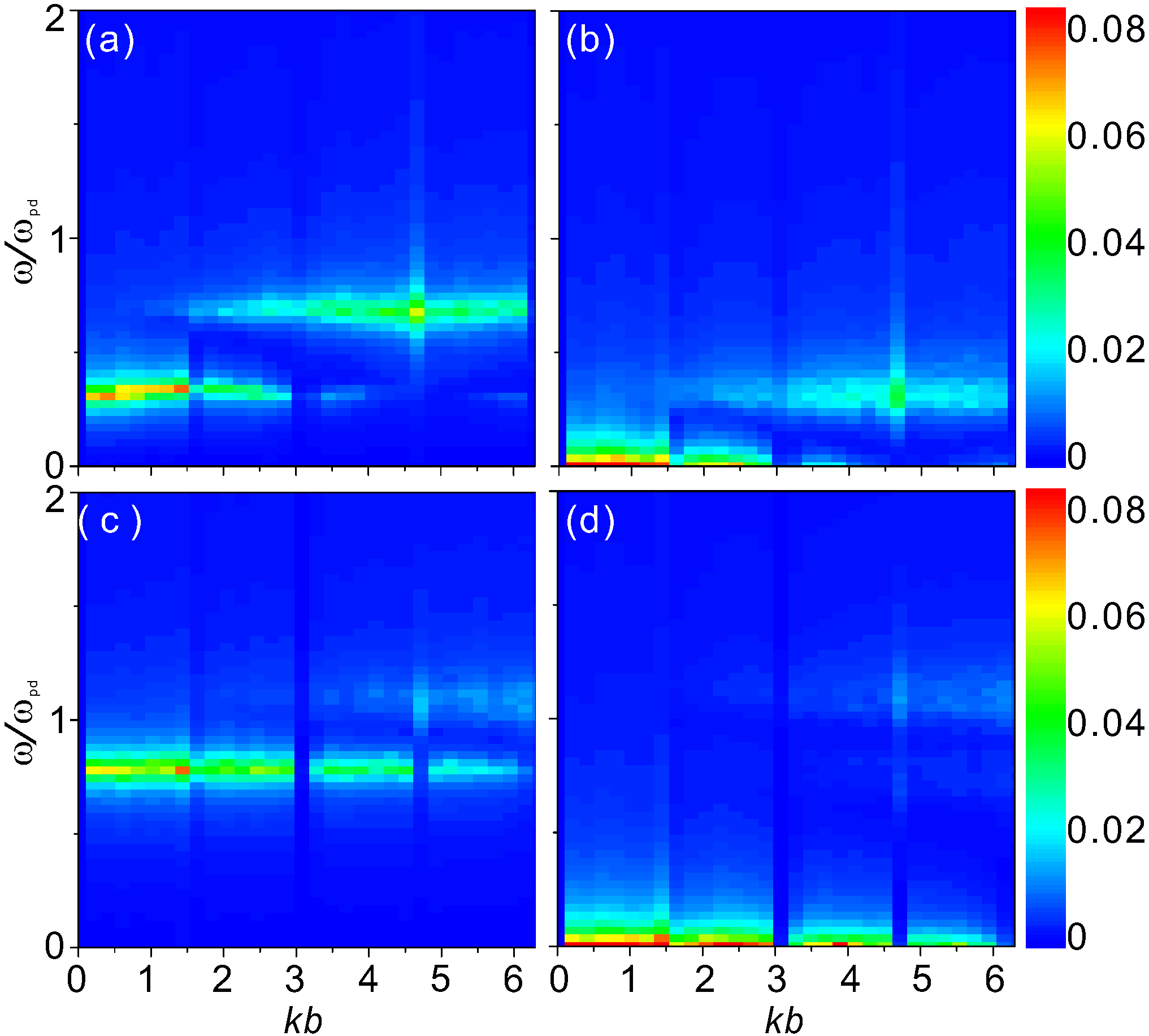}
    \caption{\label{fig:Wid2DepX}
          The longitudinal and transverse phonon spectra $\tilde{C}_{L}({\bf k}_x,\omega)$ (a, c) and $\tilde{C}_{T}({\bf k}_x,\omega)$ (b, d) as functions of $\omega/\omega_{pd}$ and $kb$ for two different periodic substrates of $U_0=0.25 E_0$ in (a, b) and $U_0 = E_0$ in (c, d), where $w=2.004b$ is unchanged.
      }
\end{figure}

We next consider the effect of changing the depth of the substrate, or the substrate strength. Figure~\ref{fig:Wid2DepX} shows $\tilde{C}_{L}({\bf k}_x,\omega)$ and $\tilde{C}_{T}({\bf k}_x,\omega)$ for
substrates with $w = 2.004b$ at
substrate depths of $U_0 = 0.25 E_0$ for
Fig.~\ref{fig:Wid2DepX}(a,b)
and $U_0=E_0$ for
Fig.~\ref{fig:Wid2DepX}(c,d). As the depth of the substrate increases, the frequencies of $\tilde{C}_{L}({\bf k}_x,\omega)$ and $\tilde{C}_{T}({\bf k}_x,\omega)$ both increase due to the larger restoring force exerted by the substrate on the dust particles. In Figs.~\ref{fig:Wid2DepX}(a) and (c), the lower branch of $\tilde{C}_{L}({\bf k}_x,\omega)$ intersects with the frequency axis at $\omega/\omega_{pd}=0.31$ and $\omega/\omega_{pd}=0.80$, respectively. These two values are
close to the sloshing frequencies of two combined dust particles inside the two substrates, which are $0.42 \omega_{pd}$ and $0.84 \omega_{pd}$, respectively.
By comparing all panels in Fig.~\ref{fig:Wid2DepX} with Fig.~\ref{fig:Wid2}(a, b), we find that as the substrate depth increases, the intensity of the upper branch in $\tilde{C}_{L}({\bf k}_x,\omega)$ and $\tilde{C}_{T}({\bf k}_x,\omega)$ gradually decreases. As shown in Fig.~\ref{fig:Distribution}, an increase in $U_0$
causes the two particle chains within each potential well to compress and gradually merge into a single chain, such as in Fig.~\ref{fig:Distribution}(d). As a result, the amount of energy in the high frequency mode diminishes.

\begin{figure}[htb]
    \centering
        \includegraphics{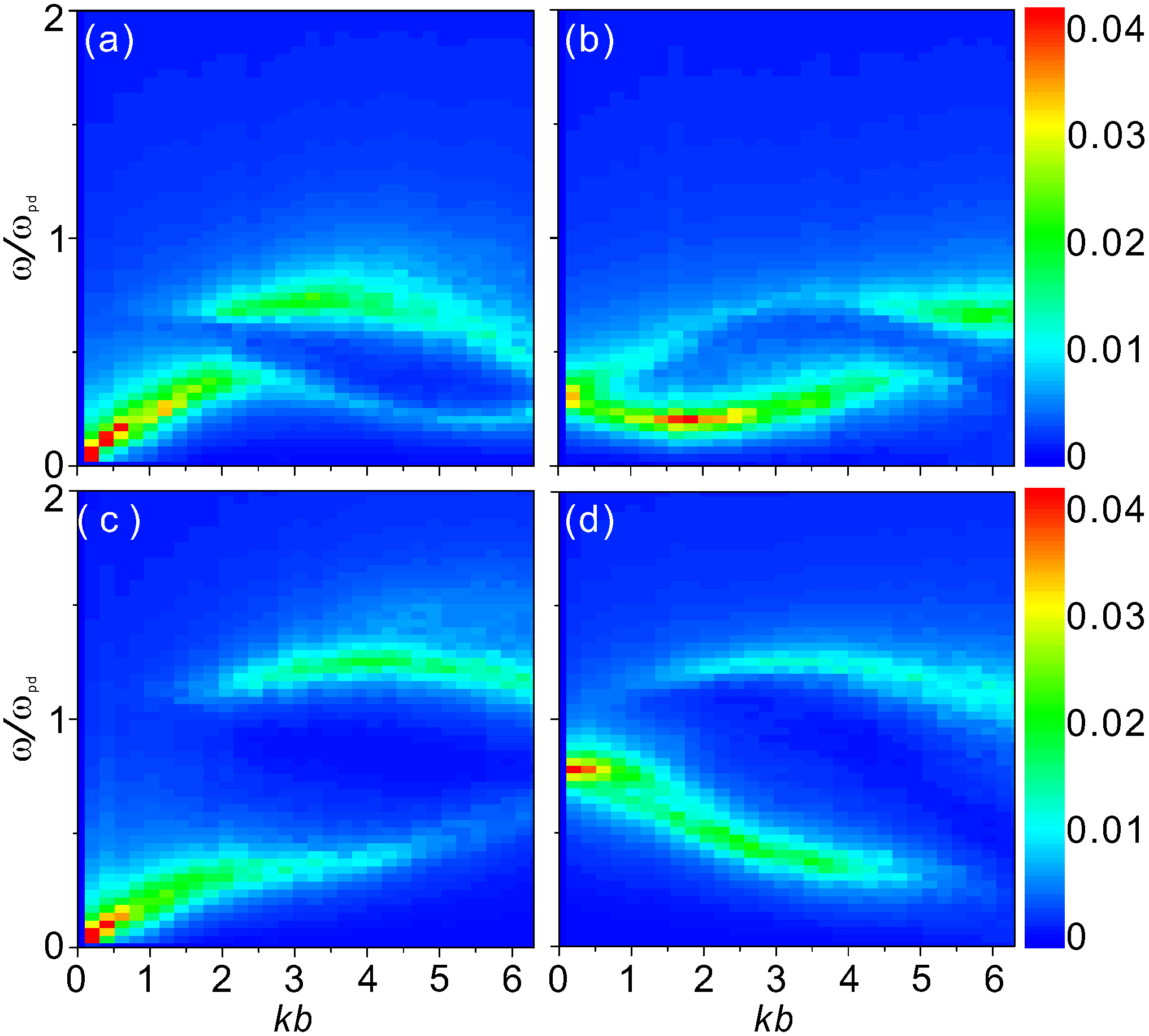}
    \caption{\label{fig:Wid2DepL}
          The longitudinal and transverse phonon spectra $\tilde{C}_{L}({\bf k}_y,\omega)$ (a, c) and $\tilde{C}_{T}({\bf k}_y,\omega)$ (b, d) as functions of $\omega/\omega_{pd}$ and $kb$ for two different periodic substrates of $U_0=0.25 E_0$ in (a, b) and $U_0 = E_0$ in (c, d), where $w=2.004b$ is unchanged.
      }
\end{figure}

Figure~\ref{fig:Wid2DepL} shows $\tilde{C}_{L}({\bf k}_y,\omega)$ and $\tilde{C}_{T}({\bf k}_y,\omega)$ spectra for the same substrates in Fig.~\ref{fig:Wid2DepX} with $w=2.004b$ at depths of $U_0=0.25 E_0$ for
Fig.~\ref{fig:Wid2DepL}(a, b) and $U_0=E_0$ for
Fig.~\ref{fig:Wid2DepL}(c,d). As the substrate depth increases,
the upper branches of $\tilde{C}_{L}({\bf k}_y,\omega)$ and $\tilde{C}_{T}({\bf k}_y,\omega)$ are both enhanced and extend to higher frequencies. This trend is exactly the same as what is shown in Fig.~\ref{fig:Wid2DepX}, and emerges for the same reason. When $kb \approx 0$, the frequencies of $\tilde{C}_{T}({\bf k}_y,\omega)$ in Fig.~\ref{fig:Wid2DepL}(b, d) are the same as those shown in Fig.~\ref{fig:Wid2DepX}(a, c), meaning that as the substrate depth increases, the frequency of $\tilde{C}_{T}({\bf k}_y,\omega)$ at $kb \approx 0$ is higher. Since this frequency is higher for a deeper substrate, the wavenumber range of the backward wave is larger, as shown in Fig.~\ref{fig:Wid2DepL}(d).

\section{Summary}

We have investigated the phonon spectra of a 2D dusty plasma confined by a one-dimensional periodic substrate. In the absence of a substrate, our results agree well with previous theoretical predictions. In the presence of the substrate, waves along the $x$ direction are unable to propagate due to the confinement of the particles by the substrate minima. For narrow or deep substrate minima, a single one-dimensional chain of particles occupies each potential well, and waves along the $y$ direction are dominated by the longitudinal motion of this single chain.  When the substrate minima are wider or shallower, the particles buckle into a zigzag structure, splitting the spectra into a sloshing branch and a breathing branch.  At small wave numbers, the sloshing wave propagates backward due to the repulsion between neighboring dust particles, and the wavenumber range of this backward motion increases as the substrate becomes deeper.  Our results could be tested experimentally by applying optically generated substrates to a dusty plasma.

We thank J.~Goree for helpful discussions. Work in China was supported by the National Natural Science Foundation of China under Grant No. 11505124, the 1000 Youth Talents Plan, the Six Talent Peaks project of Jiangsu Province, and startup funds from Soochow University. Work at LANL was carried out under the auspices of the NNSA of the U.S. DOE under Contract No. DE-AC52-06NA25396.

\end{document}